# Low-Temperature Optical Characterization of Single CdS Nanowires

L. V. Titova, Thang B. Hoang, H. E. Jackson and L.M. Smith
Department of Physics
University of Cincinnati
Cincinnati, OH

J.M. Yarrison-Rice
Physics Department
Miami University
Oxford, OH

J.L. Lensch and L.J. Lauhon
Department of Materials Science and Engineering,
Northwestern University
Evanston, IL

*Abstract*—We use spatially resolved micro-PL imaging at low temperature to study optical properties of two sets of CdS nanowires grown using 20 nm and 50 nm catalysts. We find that low temperature PL of single nanowires is an ideal technique to gauge the quality of a given growth run, and moreover enables the collection of detailed spatial information on single wire electronic states.

*Keywords-nanowires; CdS; photoluminescence imaging*

I. INTRODUCTION

It has been recently shown that CdS nanowires (NWs) can be successfully used as active elements in nanoscale optical devices, such as photodetectors, waveguides, and nanoscale lasers operating in the visible wavelength region [1-3]. However, despite significant progress in nanowire device development, detailed understanding of the effect of NW size and morphology on their optical properties remains limited.

In this paper we use direct PL micro-imaging at low temperatures to study the optical properties of two sets of CdS NWs grown using the vapor-liquid-solid (VLS) method and colloidal gold catalysts [4]. Through imaging of ~10 wires from each set, which were grown using distinct reactant delivery methods, we are able to make general comparisons of the nanowire properties resulting from these two different growth runs, and furthermore we can make detailed measurements of the electronic properties of single nanowires along the growth axis. We believe that these results point the way to rapid nondestructive testing of single nanowires which can later be utilized in the fabrication of working devices.

II. EXPERIMENTAL DETAILS

For optical measurements on individual NWs, the two sets of CdS NWs were dispersed onto silicon substrates and placed into a variable temperature continuous flow helium cryostat.

The PL measurements were conducted using slit-confocal micro-photoluminescence with 1.2 μm resolution [5]. A 50X/0.5NA long working length microscope objective was used to project a 250X magnified image of single CdS wires onto the entrance slit of the spectrometer.

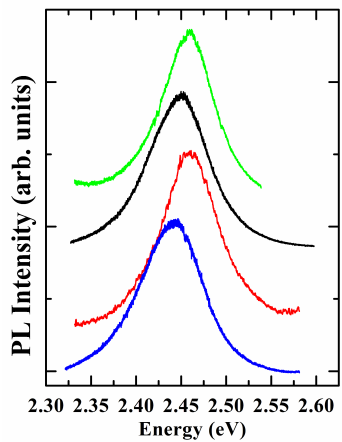

Figure 1. PL spectra of representative nanowires from the 20 nm catalyst wires (top spectrum) and 50 nm catalyst wires (other three spectra) taken at 295 K.

Several individual NWs were excited by 2.5mW of a 458 nm Ar+ laser defocused to 20 μm spotsize, and their PL emission was dispersed by a DILOR triple spectrometer working in subtractive mode, and detected by a 2000 x 800 pixel liquid nitrogen-cooled CCD detector.

In order to obtain both spectral and spatial information on single CdS nanowires, a Dove prism was used to rotate the 250X image of the nanowire so that it was oriented along the entrance slit of the spectrometer. In this configuration, a single CCD image contains both spatial information (in the direction along the spectrometer slit, or along the nanowire), as well spectral information. The spatial resolution of the optical system was 1.2 μm.

## III. RESULTS AND DISCUSSION

In this paper we show detailed results from four NWs grown using 20 nm gold catalysts and six NWs grown using 50 nm gold catalysts. The 50 nm nanowires were grown with the CdS single-source precursor placed at the upstream end of a quartz tube furnace. The growth substrate, held at 525°C, was placed downstream of the precursor, which was held at ~200°C. The 20 nm nanowires were grown at the same substrate temperature, but the precursor was heated in a separate chamber to ~230°C and delivered to the reactor with a hydrogen carrier gas flow. Independent heating of the CdS precursor resulted in better control over NW morphology. The 20 nm nanowires were straight, uniform, and 3.5 - 5 μm in length, while the 50 nm nanowires were on average three times longer (12 – 15 μm) and displayed more varied morphological variations. Because the exciton Bohr radius is only 2.8 nm in CdS, neither set of wires are expected to show any quantum confinement effects.

From Fig. 1, which displays the representative PL spectra of the 20 nm and 50 nm nanowires at 295 K, it is clear that the room temperature optical properties of all nanowires of either type are similar. At room temperature all nanowires of either type show a broad (FWHM ~ 78 meV) emission line centered near 2.45 eV. On the other hand, at low temperature significant differences between the two sets of nanowires can be seen (See Fig. 2). Each sets of wires shows two features, near-band-edge (NBE) emission from 2.52 to 2.54, and a broad defect emission band extending from 2.43 to 2.5 eV which is most likely related to defect or impurity luminescence in the wires. However, the 20 nm catalyst NWs generally show dominant NBE emission with weaker defect band emission. In contrast, the 50 nm catalyst NWs generally show defect band emission which can be as strong, or stronger than, the NBE emission. Moreover, the longer 50 nm catalyst NWs also occasionally show a number of sharp emission lines localized to the sites of morphological defects along the nanowires [6]. Remarkably, the defects and structural irregularities manifest themselves only in low temperature PL spectra, while the room temperature spectra are unaffected by them.

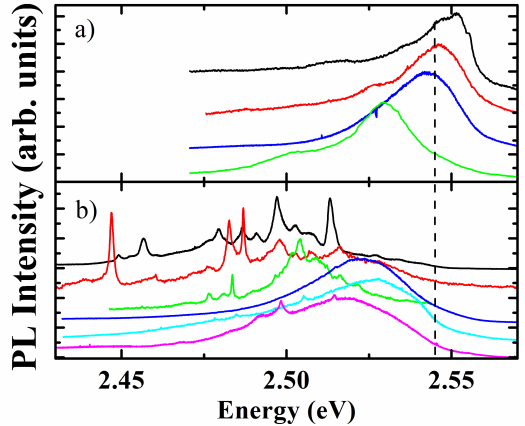

Figure 2. PL spectra of (a) set A NWs and (b) set B NWs taken at 7 K.

Through such surveys of single NWs, it becomes possible to select wires from both growths which show no defect band emission or morphological defects (sharp lines) and emit a single featureless NBE line. To gain more detailed comparisons between these selected wires, we use spatially resolved direct PL imaging of single nanowires, which has been described in detail in Reference [6].

In Fig. 3, the PL spectra extracted from the spatial PL maps of two selected wires from the 20 nm catalyst growth (Fig. 3 (a)) and the 50 nm catalyst growth (Fig. 3 (b)) at positions ~1.4 μm apart that display only the NBE emission and no measurable defect band emission. Each spectrum is normalized for clarity. While the PL of the 50 nm catalyst nanowires (Fig. 3 (b)) is free from defect band emission or morphological imperfections, the PL peak positions vary significantly (2.521± 0.007 eV) at different positions along the 14 μm length of the NW. For comparison, the spatial variation of emission energy of the 20 nm catalyst (Fig. 3 (a)) wire is significantly smaller, 2.541 ± 0.002 eV (note that the NW is only 5 μm long). It is natural to presume that these spatial variations result from morphological inhomogeneities since the 20 nm NWs are much smoother than the 50 nm NWs. On the other hand, we cannot completely rule out the possibility of non-uniform strain along the length of these nanowires. Thus, the spatially resolved PL imaging provides information about the uniformity and variations of the local energy level structure even in case of the uniform wires that do not exhibit pronounced localized defect emission.

## IV. CONCLUSIONS

In this paper we have shown that low temperature spatially resolved PL measurements of single nanowires provide a sensitive measure of NW quality. Unlike structural characterization like high-resolution SEM which can quench PL in CdS nanowires, low temperature PL imaging is completely nondestructive. Such measurements can be used to make general comparisons between the quality of a particular growth run (e.g., the relative ratio of NBE emission to defect band emission), as well as specific comparisons

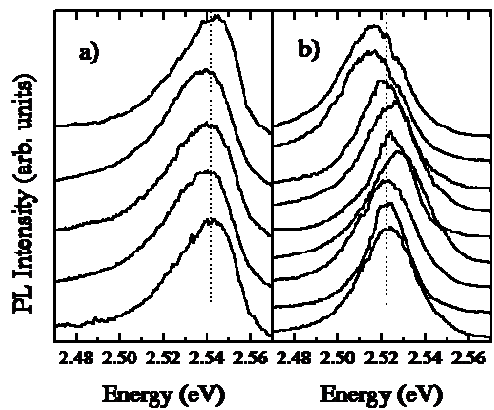

Figure 3. Spectra taken along different positions along the wires (20 nm catalyst wire (a) and 50 nm catalyst wire (b) ) with 1.4 µm step.

between two nanowires which appear to be free from defects or morphological compromises.

Comparison between two such "selected" NWs from each growth run show that NWs grown using 20 nm catalyst display significantly fewer defects than the 50 nm catalyst growths. Moreover, even in the absence of measurable defect luminescence, the 20 nm growth displays more spatially uniform emission along the length of the wire. These differences reflect the improved precursor control in the 20 nm NW growth run.

ACKNOWLEDGMENT

The authors acknowledge equipment support of the National Science Foundation through Grant Nos. DMR 0071797 and 0216374, and operating support from the University of Cincinnati. This work was partially supported by the Petroleum Research Fund of the American Chemical Society and Northwestern University. J.L.L. acknowledges the support of a National Science Foundation Graduate Research Fellowship.